\documentclass[pra,twocolumn,showpacs,preprintnumbers]{revtex4-1}
\usepackage{amsmath,amssymb,bm,graphicx,hyperref}

\renewcommand\d{\partial}
\newcommand\grad{\bm{\nabla}}
\renewcommand\r{\bm{r}}

\begin{document}
\preprint{LA-UR-11-12008}

\title{Impossibility of the Efimov effect for $p$-wave interactions}

\author{Yusuke~Nishida}
\affiliation{Theoretical Division, Los Alamos National Laboratory,
Los Alamos, New Mexico 87545, USA}

\begin{abstract}
 Whether the Efimov effect is possible, in principle, for $p$-wave or
 higher partial-wave interactions is a fundamental question.  Recently,
 there has been a claim that three nonrelativistic particles with
 resonant $p$-wave interactions exhibit the Efimov effect.  We point out
 that the assumed $p$-wave scattering amplitude inevitably causes a
 negative probability.  This indicates that the Efimov states found
 there cannot be realized in physical situations.  We also restate our
 previous argument that the Efimov effect, defined as an infinite tower
 of universal bound states characterized by discrete scale invariance,
 is impossible for $p$-wave or higher partial-wave interactions.
\end{abstract}

\date{November 2011}


\maketitle

When nonrelativistic particles attract with resonant $s$-wave
interactions, three particles always exhibit the Efimov effect, i.e., an
infinite tower of universal bound states characterized by discrete scale
invariance~\cite{Efimov:1970}.  Since the Efimov effect is one of the
most beautiful phenomena in few-body physics, it is fundamentally
important to ask whether the Efimov effect is possible, in principle,
for $p$-wave or higher partial-wave interactions.

Recently, there has been a claim that three nonrelativistic particles
with resonant $p$-wave interactions exhibit the Efimov
effect~\cite{Braaten:2011}.  Their analysis is based on the assumption
that the effective range approximation for the $p$-wave scattering
amplitude,
\begin{equation}\label{eq:amplitude}
 f_p(k,\theta) = \frac{k^2\cos\theta}{-\frac1{a_p}+\frac12r_pk^2-ik^3},
\end{equation}
is valid up to $k\to\infty$ with fixed $a_p$ and $r_p$, or more
formally, for $k<\Lambda$ with $\Lambda\gg|a_p|^{-1/3},|r_p|$.  We point
out that this assumption inevitably causes a negative probability
regardless of underlying interactions.  This indicates that their
assumption and, thus, conclusions cannot be realized in physical
situations.  Our points also apply to the earlier work in
Ref.~\cite{Macek:2006}.

\subsubsection*{Non-normalizability of the wave function}
In terms of the wave function, the use of Eq.~(\ref{eq:amplitude}) up to
$k\to\infty$ is equivalent to imposing the condition that the singular
behavior of the wave function for a small separation $r$ between two
interacting particles,
\begin{equation}
 \Psi(\r) \propto Y_1^m(\hat\r)\frac1{r^2} + \cdots,
\end{equation}
extends to $r\to0$.  Obviously, such a wave function is too singular to
be normalized.  The same is true for the three-particle wave functions
of the Efimov states claimed in Refs.~\cite{Braaten:2011,Macek:2006}.

\subsubsection*{Negative probability in the potential}
To gain further insight into the normalization issue, we regularize the
short-distance behavior by a finite-range potential that is nonzero at
$r<R$ and otherwise vanishes.  When a two-particle bound state with
binding energy $E=-\hbar^2\kappa^2/m$ exists, its normalized wave
function outside the potential range is given
by~\cite{Pricoupenko:2006,Jona-Lasinio:2008}
\begin{equation}
 \Psi(\r)\big|_{r>R} = \frac{Y_1^m(\hat\r)}{\sqrt{-\frac{r_p}2-\frac{3\kappa}2}}
  \frac{\d}{\d r}\!\left(\frac{e^{-\kappa r}}r\right).
\end{equation}
References~\cite{Pricoupenko:2006,Jona-Lasinio:2008} obtained the
constraint on the effective range parameter near the resonance
$\kappa R\simeq0$ by requiring that the probability outside the
potential should not exceed 1;
$\int_{r>R}\!d\r|\Psi(\r)|^2\simeq-2/(r_pR)\leq1$.  The same constraint
can be obtained by a different approach~\cite{Hammer:2009}.

On the other hand, if the short-range potential $R\to0$ equivalent to
the large cutoff $\Lambda\to\infty$ is considered with fixed $a_p$ and
$r_p$, the probability outside the potential exceeds 1.  As a
consequence, the normalization of the wave function is ensured only by a
negative probability in the potential; $\int_{r<R}\!d\r|\Psi(\r)|^2<0$,
which is obviously unphysical.  Similarly, in two-channel models of the
Feshbach resonance~\cite{Jona-Lasinio:2008}, the occupation probability
of the closed-channel molecule becomes negative in the same limit.
Therefore, for model interactions studied in
Refs.~\cite{Pricoupenko:2006,Jona-Lasinio:2008}, the absence of negative
probabilities unexceptionally requires the lower bound on
$|r_p|\geq O(1)/R$ near the resonance $|a_p|^{1/3}\gg R$.

\subsubsection*{Violation of the unitarity bound}
We now generally argue that the above negative probability is inevitable
regardless of underlying interactions as long as
Eq.~(\ref{eq:amplitude}) is assumed with
$\Lambda\gg|a_p|^{-1/3},|r_p|$.  Reference~\cite{Nishida:2010} showed
that scaling dimensions of local operators are bounded from below by
requiring that norms of states are non-negative.  However, for an
effective field theory incorporating
Eq.~(\ref{eq:amplitude})~\cite{Bertulani:2002},
\begin{align}\label{eq:EFT}
 \mathcal{L} &=\psi^\dagger\left(i\hbar\d_t+\frac{\hbar^2\nabla^2}{2m}\right)\psi
 - \bm\phi^\dagger\left(i\hbar\d_t+\frac{\hbar^2\nabla^2}{4m}+\nu\right)\bm\phi
 \notag \\
 &\quad + g\,\bm\phi^\dagger\cdot\psi\tensor\grad\psi
 + g\,\psi^\dagger\tensor\grad\psi^\dagger\cdot\bm\phi
\end{align}
(note the unconventional sign in front of the second term), the
two-particle operator $\bm\phi\sim\psi\tensor\grad\psi$ turns out to
have the scaling dimension $\Delta=1$, which violates the unitarity
bound $\Delta\geq3/2$~\cite{Nishida:2010}.  Therefore, the negative-norm
state must exist in the two-particle sector of the effective field
theory (\ref{eq:EFT}).

The existence of a negative-norm state or negative probability indicates
that no physical interaction can realize Eq.~(\ref{eq:amplitude}) for
$k<\Lambda$ with $\Lambda\gg|a_p|^{-1/3},|r_p|$~\cite{footnote1}.  This,
in turn, means that the scaling region
$|a_p|^{-1/3},|r_p|\ll k\ll\Lambda$, in which the $p$-wave scattering
amplitude becomes scale invariant and accordingly the Efimov states
appear~\cite{Braaten:2011}, does not exist.  Therefore, the Efimov
states claimed in Refs.~\cite{Braaten:2011,Macek:2006} are not
physically realizable bound states but mirages seen by artificially
assuming Eq.~(\ref{eq:amplitude}) with $\Lambda\gg|a_p|^{-1/3},|r_p|$
which is never achieved by physical interactions~\cite{footnote2}.
Since this is the fundamental issue, further discussion is worthwhile.

Finally, we take this opportunity to restate our argument from
Ref.~\cite{Nishida:2011} that the Efimov effect is impossible for
$p$-wave or higher partial-wave interactions.  Recall that the Efimov
effect is defined as the formation of an infinite tower of universal
bound states characterized by discrete scale invariance.  In order for
three-particle systems to exhibit the discrete scale invariance, it is
necessary that the two-particle interactions are scale invariant because
the symmetry can be broken by the quantum anomaly but cannot be
enhanced.  By setting the scattering length to be infinite and all
effective range parameters to be zero, the $\ell\,$th partial-wave
scattering amplitude becomes scale invariant.  However, as we have
pointed out here, such an interaction for any $\ell\geq1$ causes the
non-normalizability of the wave function or a negative probability if
the normalization is imposed.  Therefore, as long as we respect
principles of quantum mechanics, the Efimov effect is impossible for
$p$-wave or higher partial-wave interactions.  This observation led the
authors of Ref.~\cite{Nishida:2011} to the idea of mixed dimensions in
their search for new systems exhibiting Efimov physics.

The author thanks P.~F.~Bedaque, E.~Braaten, Y.~Castin, J.~E.~Drut,
B.~F.~Gibson, H.-W.~Hammer, D.~B.~Kaplan, D.~R.~Phillips,
L.~Pricoupenko, G.~Shen, D.~T.~Son, S.~Tan, and P.~Zhang for valuable
discussions and comments on this article and a LANL Oppenheimer
Fellowship for the support of this work.

\subsubsection*{Note added}
Recently, it was explicitly shown that the $p$-wave scattering amplitude
(\ref{eq:amplitude}), at the resonance $1/a_p=0$, contains a
contribution of the negative-norm state with binding energy
$E=-\hbar^2r_p^2/(4m)$~\cite{Braaten:2012}.  This means that the
validity of Eq.~(\ref{eq:amplitude}) has to be taken as $k<\Lambda$ with
$\Lambda\ll|r_p|$, which is consistent with the conclusion of this
article.

\end{document}